\documentclass[journal]{IEEEtran}

\usepackage{cite}

\ifCLASSINFOpdf
\else
\fi

\usepackage{graphicx}
\usepackage{amsmath}
\usepackage{mathtools}
\usepackage{commath}
\usepackage{comment}
\usepackage{enumerate}

\usepackage{booktabs}
\usepackage{multirow} 

\usepackage{epstopdf}
\usepackage{array}
\usepackage{wrapfig}

\usepackage{subcaption}

\newcommand{\innermid}{\nonscript\;\delimsize\vert\nonscript\;}
\newcommand{\activatebar}{%
  \begingroup\lccode`\~=`\|
  \lowercase{\endgroup\let~}\innermid 
  \mathcode`|=\string"8000
}

\usepackage{amsthm,amssymb}
\usepackage{setspace}

\newcommand{\subparagraph}{}
\usepackage[compact]{titlesec}

\usepackage{url}
\makeatletter
\g@addto@macro{\UrlBreaks}{\UrlOrds}
\makeatother

\usepackage{enumerate}

\begin{document}

\bstctlcite{IEEEexample:BSTcontrol}


%
\title{An Augmented Autoregressive Approach to HTTP Video Stream Quality Prediction}
%
%
%

\author{Christos G. Bampis and Alan C. Bovik
\thanks{C. G. Bampis and A. C. Bovik are with the Department
of Electrical and Computer Engineering, University of Texas at Austin, Austin,
USA (e-mail: bampis@utexas.edu; bovik@ece.utexas.edu). This work is supported by Netflix Inc.}
}

\maketitle

\begin{abstract}

HTTP-based video streaming technologies allow for flexible rate selection strategies that account for time-varying network conditions. Such rate changes may adversely affect the user's Quality of Experience; hence online prediction of the time-varying subjective quality can lead to perceptually optimized bitrate allocation policies. Recent studies have proposed to use dynamic network approaches for continuous-time prediction; yet they do not consider multiple video quality models as inputs nor consider forecasting ensembles. Here we address the problem of predicting continuous-time subjective quality using multiple inputs fed to a non-linear autoregressive network. By considering multiple network configurations and by applying simple averaging forecasting techniques, we are able to considerably improve prediction performance and decrease forecasting errors. 

\end{abstract}

\begin{IEEEkeywords}
HTTP video quality, NARX, Quality of Experience prediction
\end{IEEEkeywords}

\IEEEpeerreviewmaketitle

\section{Introduction}

Streaming video data now occupies a considerable fraction of available wireline and wireless network \cite{cisco} bandwidth, which locally, can be highly volatile. To address the time-varying properties of the available bandwidth, HTTP-based video technologies have been recently introduced, thereby allowing for adaptations to the client's playout rate \cite{chen2015rate, huang2015buffer}. Streaming content providers such as Netflix and Youtube are applying adaptive bitrate selection schemes as they seek to reduce the number of playback interruptions, also known as rebuffering events, towards maximizing the users' Quality of Experience (QoE). However, accurately predicting subjective QoE is subject to various difficulties, largely arising from the non-linear processes found along the visual pathway and the non-linearities inherent to human subjective responses, particularly when there are temporo-behavioral aspects involved.

There have been numerous approaches taken to QoE prediction. In video streaming applications, there are two major video impairments: compression artifacts resulting from rate adaptations, and playback interruptions due to buffer outage. Broadly, video quality assessment (VQA) models are designed to predict the perceptual effects of video impairments during normal video playback, i.e., when no rebuffering occurs, while Quality of Service (QoS) approaches focus on modeling the effects of playback interruptions on user QoE. Both VQA and QoS models can be used to conduct either retrospective (computing a single number reflecting the overall QoE) or continuous-time (instantaneous) QoE. Continuous-time QoE prediction is fundamentally different from retrospective prediction, given the time-dependencies that arise when addressing continuous-recored subjective quality.

Video quality assessment is a large, intensively-studied research area, encompassing full-reference models (FR), when reference information is available, (e.g., SSIM \cite{wang2004video} and MOVIE \cite{seshadrinathan2010motion}), reduced reference models (RR) such as the ST-RRED indices \cite{soundararajan2013video}, and no-reference methods (NR), which are relevant when only the distorted video is available, such as \cite{mittal2016completely, bovik2013automatic, moorthy2011visual, saad2014blind}. VQA models rely on models of temporal visual perception and on models of statistical regularities inherent to natural videos to quantify the severity of video distortions. Approaches to the problem of continuous-time QoE prediction have used the Hammerstein-Wiener (HW) model \cite{chen2014modeling} and a non-linear autoregressive network (NARX) approach \cite{7931662, lin1996learning}. Unlike VQA, QoS models rely on rebuffering-aware statistics, such as the number, location and duration of rebuffering events \cite{hossfeld2011quantification, rodriguez2012quality}. For example, a HW model was proposed that predicts the continuous-time effects of rebuffering on QoE in \cite{ghadiyaram2015time}. Beyond VQA and QoS models, other researchers have integrated multiple sources of information to predict QoE on videos afflicted by both rebuffering and compression, e.g. the SQI model in \cite{duanmu2016sqi} and the learning-based approaches in \cite{7931662,VATL}.

Here we focus on continuous-time QoE prediction on videos afflicted by compression artifacts. We hope to go beyond previous efforts, that suffer from two important shortcomings:
\begin{enumerate}
\item{they use only a single input measurement to predict the effects of video quality degradation on QoE}
\item{only the best (for some evaluation metric) possible time series prediction is used}
\end{enumerate}
As we will demonstrate, these problems can adversely affect the predictive power of a continuous-time QoE predictor; hence we propose an augmented NARX approach to this problem that employs multiple QoE-descriptive inputs. In a recent study \cite{7931662}, a NARX model was developed for QoE prediction. However, the work presented here is conceptually different since it focuses on improving QoE prediction on video streams afflicted by rate changes, by including multiple VQA prediction streams from different models as inputs, and by deploying a simple forecast ensemble \cite{zhang2001time, zhang2003time}. Notably, the study in \cite{7931662} showed that when applied to those video streams in the LIVE-NFLX Video QoE Database \cite{DB_paper}, that are impaired only by compression (viz., no rebuffering events), the NARX approach delivered worse performance. The augmented approach described here yields considerably better results than the original NARX approach. 

The rest of this paper is organized as follows: Section \ref{previous_works} discusses the shortcomings of previous approaches and motivates the need for an augmented approach to continuous-time QoE prediction. Then, Section \ref{NARX_again} briefly revisits the NARX model, while Section \ref{multi_inp_ensemble} describes the various inputs we selected and the details of the forecasting ensemble approach. Experimental results are given in Section \ref{experiments}, while Section \ref{the_end} concludes with a discussion of future work.

\section{Previous work on continuous-time QoE prediction}
\label{previous_works}

A common drawback of the methods proposed in \cite{chen2014modeling} and \cite{7931662} is that they do not consider more than one quality-aware input to their prediction engines. Despite the existence of various video quality models, there is no guarantee as to whether these models are sufficient to drive large and accurate QoE prediction engines, especially since simple VQA models perform poorly, and while better, more advanced ones are computationally demanding. In streaming applications, making online predictions is highly desirable; hence accuracy may have to be sacrificed to match time constraints. Further, the design of successful NR VQA models is still a largely unsolved problem, hampering their potential use in QoE applications. To sum up, the development of a single, efficient and accurate VQA model has proved to be challenging.

The second most important shortcoming of continuous-time predictors \cite{7931662, chen2014modeling, ghadiyaram2015time} is that they search for the ``best candidate" time-series prediction, usually by training and testing on different data splits, and by determining each learning engine's parameters. In our experiments, we have found that both the HW and the NARX models do not always produce stable results. On the one hand, NARX may produce different results due to different neural network initializations, while realization of the Hammerstein-Wiener may also produce unstable results. Selecting a single best-performing configuration using either NARX or HW depends heavily on the type of evaluation metric that is used. For example, the root-mean-squared-error (RMSE) may be a poor choice. More importantly, the set of model parameters (e.g. the number of NARX input lags and the HW model order) is not unique. As a result, settling on a single continuous-time QoE predictor can yield less accurate and less robust prediction forecasts \cite{zhang2001time}.

\section{Non-linear Autoregressive Approach}
\label{NARX_again}

As shown in \cite{7931662}, subjective QoE can be modelled as a non-linear aggregate of various QoE-aware inputs: predicted video quality during normal playback, and rebuffering inputs which account for the effects of rebuffering and memory-driven continuous-time features. Given the superiority of NARX over HW on this problem \cite{7931662}, we relied on the NARX model here too. We now briefly revisit the main idea behind NARX. This model assumes a non-linear relationship between the output $y_t$ at time $t$ and the inputs to the model, which are delayed versions of the output, i.e., ${y_{t-1}, \ y_{t-2}, \ y_{t-3}, ..., y_{t-d_{y}}}$, and the ``exogenous" variables at time $t$ ($\mathbf{u}_{t}$) and in the past ($\mathbf{u}_{t-1}, \mathbf{u}_{t-2}, ..., \mathbf{u}_{t-d_u}$). The autoregressive portion of the NARX model allows it to model QoE memory effects between current and past outputs, while the exogenous variables $\mathbf{u}_{t}$ contain QoE-related information (such as video quality) that affect QoE. The exogenous variables at time $t$ can be either scalars or vector-valued, depending on the number of exogenous inputs/variables that are included in the model.

\begin{figure}[htp]
\centerline{\includegraphics[width=0.85\columnwidth] {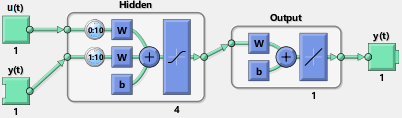}}

\hspace{0.7mm}\centerline{\includegraphics[width=0.85\columnwidth] {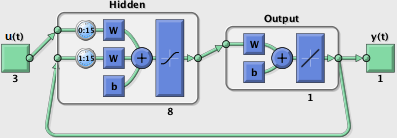}
}
\caption{Top: Open-loop (OL); Bottom: Closed-loop (CL) configurations when training NARX.}
\label{configs_demo}
\end{figure}

The non-linearity is approximated by a feed-forward neural network (see Fig. \ref{configs_demo}). The NARX architecture also allows for versatility in selecting the number of input and feedback lags $d_u$ and $d_y$ as well as the number of hidden layers and the number of neurons in each layer. There are two main options when training a NARX model: training in an open-loop (OL) configuration, where the ground truth subjective QoE scores are fed into the network when learning (as in Fig. \ref{configs_demo}), or in a closed-loop (CL) configuration, where the output prediction is used instead. Given the unavailability of subjective scores when testing or applying the NARX model, the OL configuration is only possible when training the model; the CL configuration must be used when testing or in application \cite{matlab_link}.

As shown in the experimental section, the CL configuration requires more compute time and performs worse than the OL configuration \cite{matlab_link}. However, using the augmented VQA inputs yields better results when using the CL configuration, as does the averaging ensemble. For simplicity, we deployed a NARX model using a single hidden layer similar to \cite{7931662}.

\section{Augmented Approach for QoE Prediction}
\label{multi_inp_ensemble}

\subsection{Combining VQA models}

\begin{figure*}
    \begin{subfigure}{0.66\columnwidth}
        \centering
\includegraphics[width=\columnwidth] {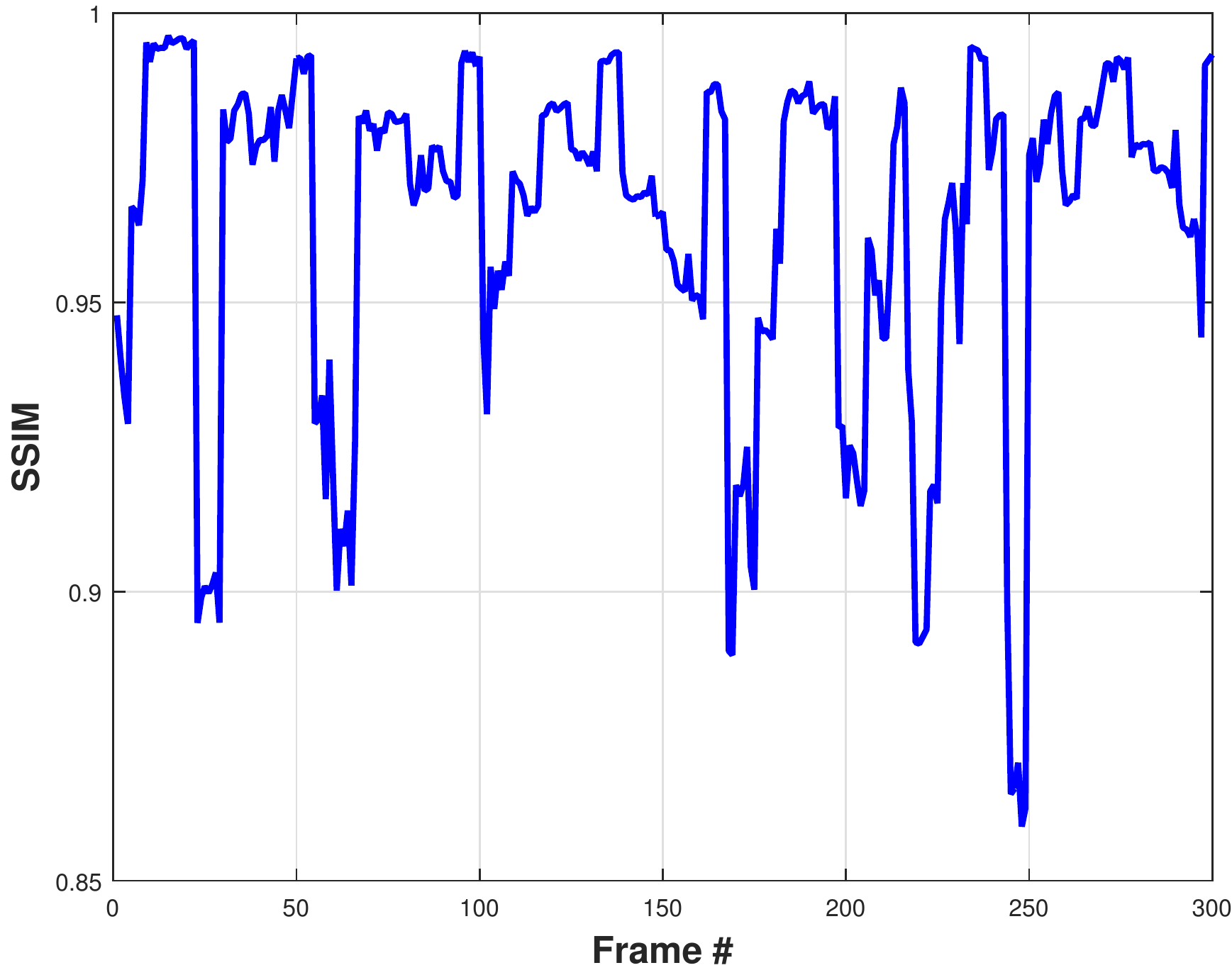}
        \caption{SSIM \cite{wang2004image} per frame}
    \end{subfigure}
    \begin{subfigure}{0.66\columnwidth}
        \centering
\includegraphics[width=\columnwidth] {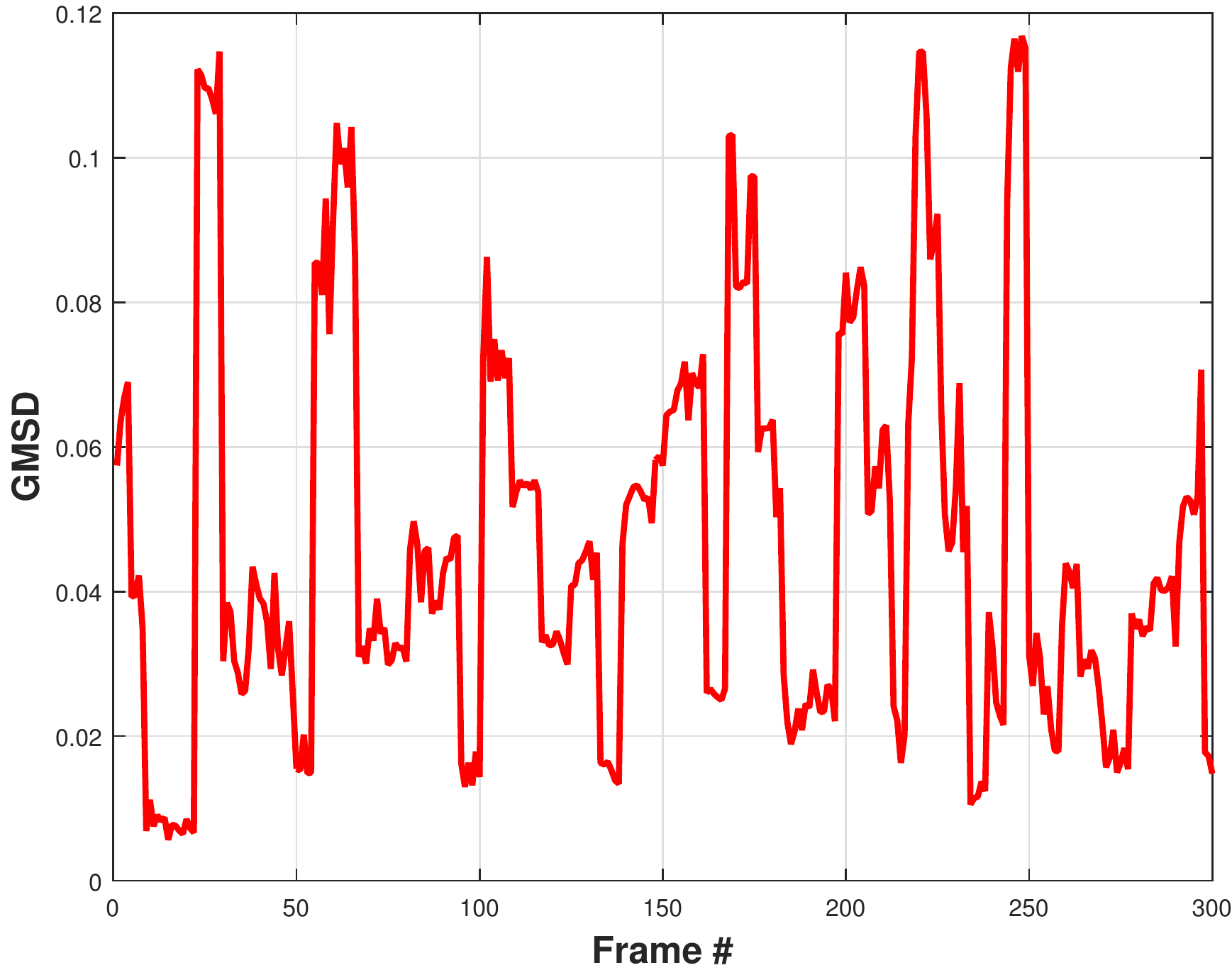}
        \caption{GMSD \cite{xue2014gradient} per frame}
    \end{subfigure}
    \begin{subfigure}{0.66\columnwidth}
        \centering
\includegraphics[width=\columnwidth] {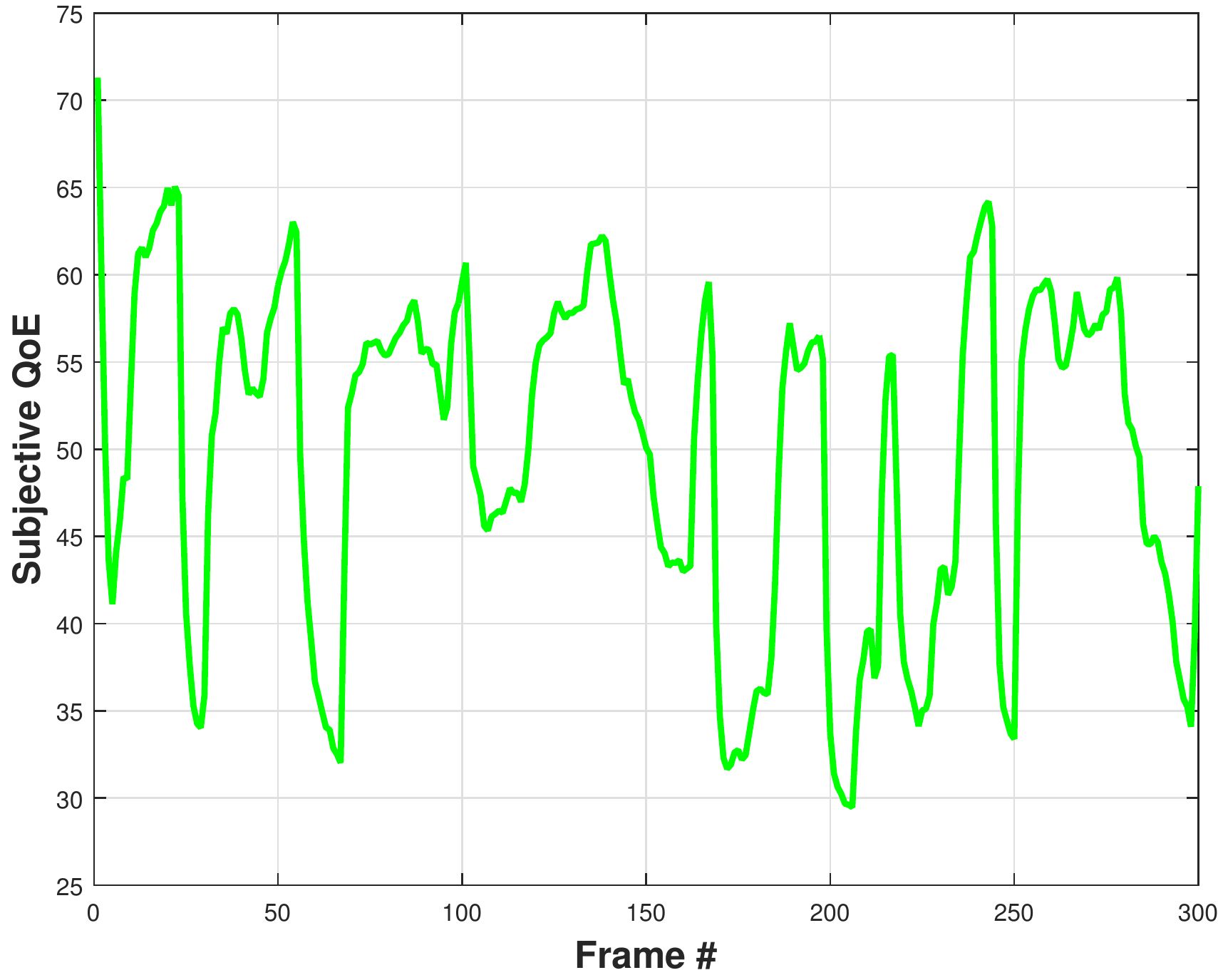}
        \caption{subjective scores}
    \end{subfigure}
    \caption{From left to right: (a) SSIM \cite{wang2004image}, (b) GMSD \cite{xue2014gradient} and (c) subjective scores on a given video sequence, drawn from the LIVE QoE Database for HTTP-based Video Streaming \cite{chen2014modeling}.}
\label{inputs_demo}
\end{figure*}

\begin{figure*}[htp]
\centerline{
\includegraphics[height = 6.7cm, width=\columnwidth] {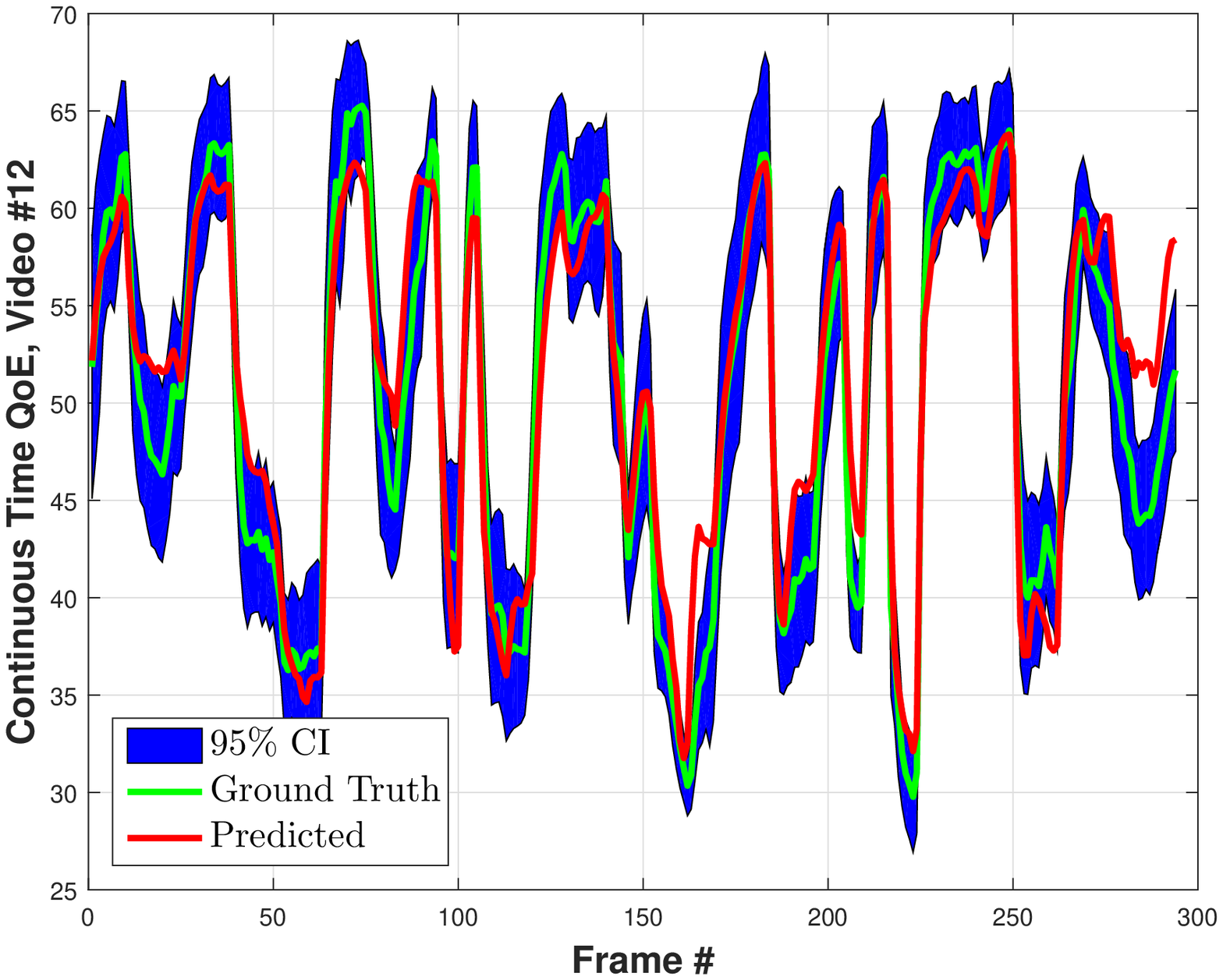}
\includegraphics[height = 6.7cm, width=\columnwidth] {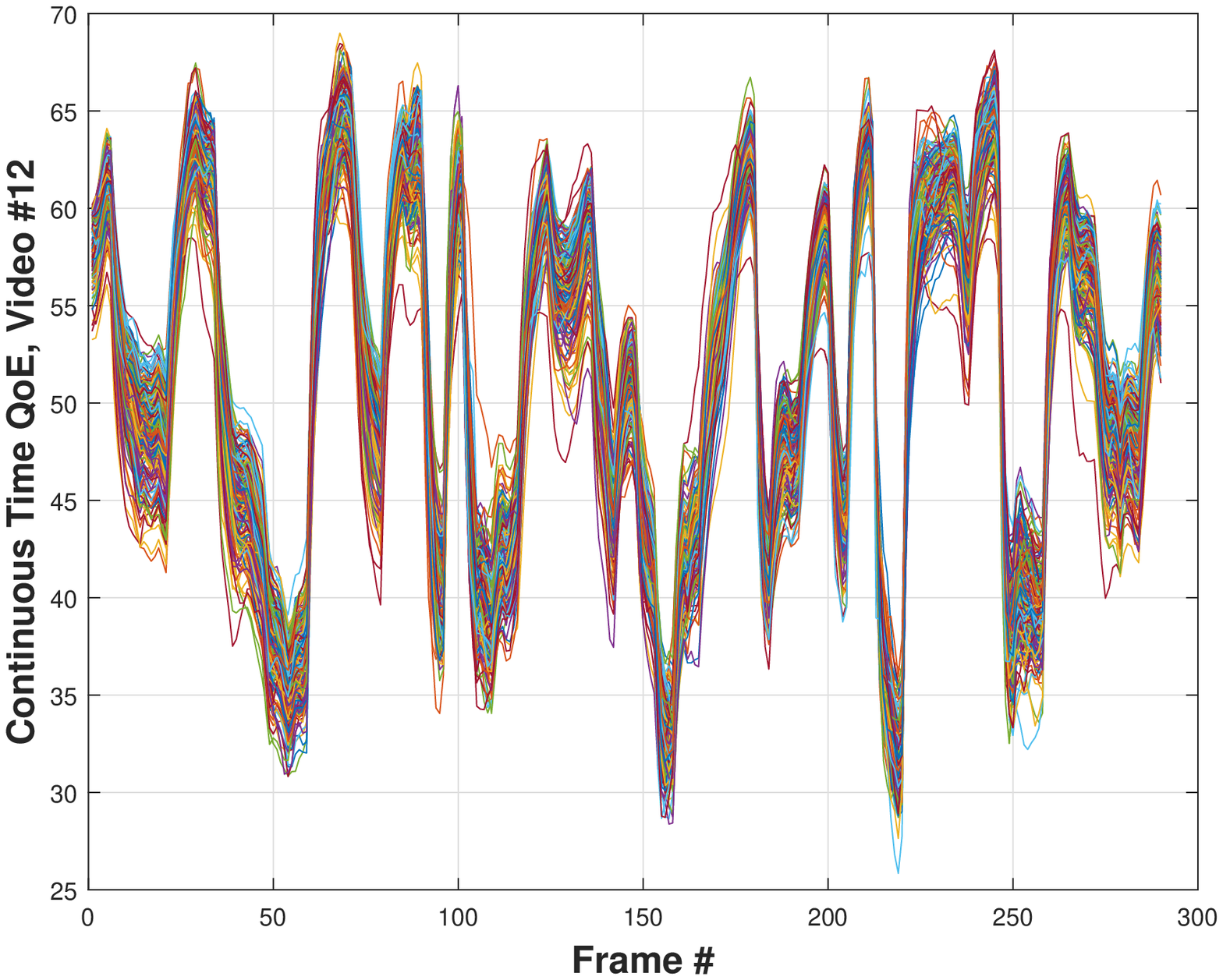}
}
\caption{a) Augmented NARX output on a sample video using ST-RRED, GMSD and SSIM. b) Plots of all the time series QoE predictions on the same video as (a).}
\label{result_pic}
\end{figure*}

We now discuss the first component of our augmented approach: combining VQA models. While this seems straightforward, increasing the number of VQA inputs can yield significant improvements in the prediction results. VQA models are not perfect, and do not always account for the same temporal distortions. By using different VQA models, we are able to reduce dependencies on each ``elementary" VQA model, thereby instead focusing their collective contributions towards better modeling video quality distortions. This approach may be thought of as a continuous-time counterpart of FVQA \cite{lin2014fusion} which later evolved into VMAF \cite{aaron2015challenges, techblog}, which is a video quality metric broadly used by Netflix. As mentioned earlier, VQA model complexity is an important factor when designing online continuous-time prediction systems. While frame-based approaches such as SSIM and PSNR are useful tools for performing simple and online VQA calculations, their performances, when these models are used in isolation, may not be sufficient on the more complex QoE problem. However, it is possible to perform simple VQA computations in parallel, rather than computing advanced (and computationally demanding) VQA models like MOVIE \cite{seshadrinathan2010motion}. This makes it possible to define VQA-augmented inputs that significantly improve QoE predictions. Likewise, NR VQA applications, where there is significant room for improvement, such an approach could also be of significant value. A variety of VQA models may be used as continuous-time inputs to our system, as shown in Fig. \ref{inputs_demo}. An example of the output of such an augmented NARX model is shown in Fig. \ref{result_pic}a. Clearly, the proposed augmented approach is able to closely track the subjective QoE.

\subsection{Averaging Forecasts}

Given a set of network configurations, e.g. by using different values of $d_u$ and $d_y$ or number of neurons in the hidden layer, a model selection scheme is needed. However, even given a model, or a model order, yielding a system that is optimal on a given set of data, it may occur that using a different number of lags (or model order) may prove more suitable for a different set of data. Meanwhile, different network initializations (given a network architecture) can produce different results when testing. Therefore, we propose to apply a simple, yet powerful, averaging of multiple different forecasts. This makes it possible to capture a wider range of temporal dependencies, by including different input and feedback lags. More importantly, more advanced ensemble forecasting methods can be employed \cite{weijs2013information}. Such ensemble techniques can be used to successfully filter out ``spurious" forecasts and to reduce the variances of the predicted forecasts. Figure \ref{result_pic}b shows the several different predictors applied on the same video. While each individual forecast captures the overall subjective quality to some extent, there can be differences in the prediction result; hence combining forecasts produces a more stable output.

\section{Experimental Results}
\label{experiments}

\begin{table}[htp]
\caption{Individual and combined VQA inputs to the NARX model. The best results per evaluation metric are denoted by boldface.}
\centering
\scalebox{1}{
\begin{tabular}{| c | c | c | c |}
    \hline
    Model/Metric & OR & SROCC & PLCC \\ \hline
    PSNR & 24.1096 & 0.7591 & 0.7452 \\ \hline
    SSIM & 4.4828 & 0.9163 & 0.9345 \\ \hline
    MS-SSIM & 5.9459 & 0.9001 & 0.9206 \\ \hline
    NIQE & 35.7931 & 0.5249 & 0.5225 \\ \hline 
    ST-RRED & 5.2740 & 0.9113 & 0.9132 \\ \hline
    GMSD & 6.6667 & 0.8976 & 0.9157 \\ \hline
    NIQE+PSNR & 8.5616 & 0.8912 & 0.8916 \\ \hline
    ST-RRED+SSIM & 3.9041 & 0.9210 & 0.9307 \\ \hline
    SSIM+MS-SSIM & 5.0000 & 0.8973 & 0.9203 \\ \hline 
    ST-RRED+GMSD+SSIM & \textbf{3.5135} & \textbf{0.9321} & \textbf{0.9396} \\ \hline
    \end{tabular}}
\label{VQA_OL}
\end{table}

We focus on the highly practical problem of continuous-time QoE prediction on video streams afflicted by HTTP-based rate changes. In our experiments, we used the LIVE QoE Database for HTTP-based Video Streaming \cite{chen2014modeling} which contains 15 video sequences, each 5 minutes long. The database consists of 3 reference videos, created by joining 8 high quality, uncompressed 720p video clips having frame rates of 30 fps. The videos were encoded using an H.264 encoder at various bitrate levels. We did not use the LIVE-NFLX Video QoE Database \cite{DB_paper}, since in this work we focus on rate changes rather than on combinations of rebuffering and dynamic rate drops. 

Following \cite{chen2014modeling}, we split the data into 10 training videos (2 reference videos) and 5 (1 reference video) testing videos. Given a set of training videos, we determined the ``best" network configuration (in terms of RMSE) by searching across various configurations: input lags $d_u \in [4 \ 10]$ with step 2, $d_y \in [4 \ 10]$ with step 2, and number of hidden neurons $5$, $8$ and $10$. We used the System Identification Toolbox in Matlab and the Levenberg-Marquardt algorithm to train NARX. After training NARX, testing was repeated 5 times to account for different network initializations. The average performance is reported. To be able to combine the predictions across all network configurations and network initializations, we also trained and tested the NARX model with each of these configurations individually. We used a wide variety of VQA models: PSNR (FR), SSIM \cite{wang2004image} (FR), MS-SSIM \cite{wang2003multiscale} (FR), NIQE \cite{mittal2013making} (NR), GMSD \cite{xue2014gradient} (FR), VMAF \cite{techblog} (FR), and ST-RRED \cite{soundararajan2013video} (RR). To compare between the predicted output and the ground truth, we used the following evaluation metrics \cite{chen2014modeling}: Spearman's rank order correlation coefficient (SROCC), Pearson's linear correlation coefficient (PLCC) and the outage rate (OR).

First, we examined the effects of combining VQA models when using the OL configuration. As shown in Table \ref{VQA_OL}, combining PSNR with NIQE delivered very large improvements across all three metrics. Similarly, combinations of more advanced pairs of VQA models, such as ST-RRED and SSIM, also improved on the prediction results. Combining three inputs (ST-RRED, GMSD and SSIM) further boosted performance. However, when using VQA models that capture similar information, such as SSIM and MS-SSIM, the predictions may not improve. Overall, the improvements will depend on whether the combined VQA inputs utilize complementary VQA-related information.

Next we considered a larger set of experiments, where we compared between the OL and CL configurations, and using or not using averaging of the time series predictions (see Table \ref{ALL_RESULTS}). Clearly, the CL approach delivered worse results than the OL configuration, across various VQA models and combinations. As with the OL results shown in Table \ref{VQA_OL}, the CL results also demonstrate the benefits of augmenting the VQA inputs. However, when using ST-RRED which deploys powerful temporal statistical image models \cite{sheikh2006image, sheikh2005visual}, the CL results were much better than for either SSIM and GMSD; hence combining either of those VQA models with ST-RRED did not improve the prediction results. Most importantly, the effect of the averaging ensemble is clear when using both the OL and CL approaches, across all VQA combinations. As an example, the results using GMSD and VMAF in the CL case, or NIQE in the OL case, were greatly improved. Note that the improvements gained by increasing the number of VQA inputs in the CL case when using ST-RRED became quite evident when using the averaging ensemble. The best combination that we tested used the OL configuration, together with the averaging ensemble, and by combining three inputs: ST-RRED, GMSD and SSIM.

\begin{table*}[htp]
\caption{Median value of OR, SROCC and LCC for the OL and CL configurations using a single best predictor (``best") or an average ensemble of the forecasts (``avg"). Best result (per evaluation metric) is denoted in boldface. The LIVE QoE Database for HTTP-based Video Streaming \cite{chen2014modeling} was used in these experiments.}
\centering
\scalebox{0.8425}{
\begin{tabular}{| c | c | c | c | c | c | c | c | c | c | c | c | c |}

    \hline
    Configuration & \multicolumn{6}{c|}{OL} & \multicolumn{6}{c|}{CL} \\ \hline
    Predictor & \multicolumn{3}{c|}{best} & \multicolumn{3}{c|}{avg} & \multicolumn{3}{c|}{best} & \multicolumn{3}{c|}{avg} \\ \hline
    
    Model/Metric & OR & SROCC & PLCC & OR & SROCC & PLCC & OR & SROCC & PLCC & OR & SROCC & PLCC  \\ \hline

    PSNR (1) & 24.1096 & 0.7591 & 0.7452 & 22.0690 & 0.7775 & 0.7676 & 27.1233 & 0.6629 & 0.6647 & 25.5172 & 0.7195 & 0.7300 \\ \hline
    SSIM \cite{wang2004image} (2) & 4.4828 & 0.9163 & 0.9345 & 4.1379 & 0.9118 & 0.9391 & 9.5205 & 0.8647 & 0.8860 & 6.2069 & 0.9056 & 0.9185 \\ \hline
    MS-SSIM \cite{wang2003multiscale} (3) & 5.9459 & 0.9001 & 0.9206 & 3.7931 & 0.9113 & 0.9276 & 10.4795 & 0.8616 & 0.8632 & 9.3103 & 0.8999 & 0.9073 \\ \hline
    NIQE \cite{mittal2013making} (4) & 35.7931 & 0.5249 & 0.5225 & 30.0000 & 0.5974 & 0.6091 & 37.0345 & 0.4474 & 0.4712 & 36.5517 & 0.5604 & 0.5580 \\ \hline
    GMSD \cite{mittal2013making} (5) & 6.6667 & 0.8976 & 0.9157 & 5.5172 & 0.9059 & 0.9228 & 12.5000 & 0.8555 & 0.8504 & 6.2069 & 0.8959 & 0.9062 \\ \hline
    VMAF \cite{techblog} (6) & 15.7823 & 0.8901 & 0.8684 & 11.7241 & 0.8936 & 0.8687 & 18.0952 & 0.7882 & 0.7852 & 11.3793 & 0.8810 & 0.8525 \\ \hline
    ST-RRED \cite{soundararajan2013video} (7) & 5.2740 & 0.9113 & 0.9132 & 4.8276 & 0.9084 & 0.9198 & 6.8027 & 0.8948 & 0.8951 & 7.2414 & 0.9086 & 0.9140 \\ \hline
    (1) + (4) & 8.5616 & 0.8912 & 0.8916 & 7.2414 & 0.9022 & 0.9132 & 15.2381 & 0.8314 & 0.8374 & 8.9655 & 0.8965 & 0.9065 \\ \hline
    (2) + (7) & 3.9041 & 0.9210 & 0.9307 & \textbf{3.4483} & 0.9224 & 0.9368 & 7.8082 & 0.8993 & 0.8975 & 5.5172 & 0.9241 & 0.9317 \\ \hline
    (2) + (3) & 5.0000 & 0.8973 & 0.9203 & 4.1379 & 0.9074 & 0.9270 & 9.4521 & 0.8846 & 0.8828 & 8.6207 & 0.9017 & 0.9092 \\ \hline
    (2) + (5) + (7) & 3.5135 & 0.9321 & 0.9396 & \textbf{3.4483} & \textbf{0.9343} & \textbf{0.9422} & 7.3288 & 0.8939 & 0.8995 & 5.1724 & 0.9189 & 0.9294 \\ \hline
    
    \end{tabular}}
\label{ALL_RESULTS}
\end{table*}

As already mentioned, not only did the CL configuration perform worse than its OL counterpart, but the training time also increased \cite{matlab_link}. Table \ref{exec_times} shows the training times required for OL and CL. For comparability, we trained a NARX model 50 times using 10 neurons, using a variable number of VQA inputs and using $d_u=d_y=10$. Then, we averaged the compute times across all of the network initializations. Clearly, the compute time increases as the number of VQA inputs increases, while training a CL NARX model required more time in all cases.
\vspace{2mm}
\begin{table}[htp]
\caption{Compute time for the OL and CL configurations (in seconds) averaged across 50 network training/testing trials. The OL configuration is less time consuming.}
\centering
\scalebox{1}{
\begin{tabular}{| c | c | c |}
    \hline
    \# inputs & OL & CL \\ \hline
    1 & 0.78 & 1.60 \\ \hline
    2 & 0.98 & 2.12 \\ \hline
    3 & 1.24 & 2.50 \\ \hline
    4 & 1.64 & 3.20 \\ \hline 
    \end{tabular}}
\label{exec_times}
\end{table}

\section{Future Work}
\label{the_end}

An augmented approach for continuous-time QoE prediction was proposed by increasing the number of VQA inputs and creating a forecasting ensemble. Both components of this approach delivered promising results. It would be very interesting to study similar approaches using other dynamic systems such as the HW model. Given the general structures of both NARX and HW, similar ideas could also be employed for QoE prediction on videos afflicted by playback interruptions. Meanwhile, more advanced ensemble forecasting approaches that assign non-equal weights to each predictor might deliver even better results. Clearly, an extensive evaluation of the proposed approach on other subjective QoE databases \cite{DB_paper} would be a great step forward in this direction.

\section{Acknowledgement}The authors thank Zhi Li, Anush Moorthy, Ioannis Katsavounidis and Anne Aaron for supporting this work.

\bibliographystyle{IEEEtran}
\bibliography{bibfile}{}

\end{document}